\documentclass[nopacs,twocolumn,aip,jcp,superscriptaddress,
letterpaper,reprint]{revtex4-2}

\pdfoutput=1

\usepackage{graphicx,bm}
\usepackage[caption=false]{subfig}
\usepackage{amsmath}
\usepackage{color}
\usepackage{hyperref}

\makeatletter
\def\fourvdots{\vbox{\baselineskip3\p@ \lineskiplimit\z@
  \kern6\p@\hbox{.}\hbox{.}\hbox{.}\hbox{.}}}
\makeatother

\usepackage[utf8]{inputenc}
\newcommand{\am}[1]{\textcolor{black}{ #1}}

\begin{document}

\begin{widetext}
\noindent\textit{The following article has been submitted to/accepted by The Journal of Chemical Physics. After it is published, it will be found at \url{https://aip.scitation.org/journal/jcp}.}
\end{widetext}

\vspace{1.cm}


\title{
Stimuli-responsive twist actuators made from soft elastic composite materials --- linking mesoscopic and macroscopic descriptions
}

\author{Andreas M. Menzel}
\affiliation{Institut f\"ur Physik, Otto-von-Guericke-Universit\"at Magdeburg, Universit\"atsplatz 2, 39106 Magdeburg, Germany}

\date{\today}

\begin{abstract}
Very recently, the construction of twist actuators from magnetorheological gels and elastomers has been suggested. These materials consist of magnetizable colloidal particles embedded in a soft elastic polymeric environment. The twist actuation is enabled by a net chirality of the internal particle arrangement. Upon magnetization by a homogeneous external magnetic field, the systems feature an overall torsional deformation around the magnetization direction. Starting from a discrete minimal mesoscopic model set-up we work towards a macroscopic characterization. The two scales are linked by identifying expressions for the macroscopic system parameters as functions of the mesoscopic model parameters. In this way, the observed behavior of a macroscopic system can in principle be mapped to and illustratively be understood from an appropriate mesoscopic picture. Our results apply equally well to corresponding soft electrorheological gels and elastomers. 
\end{abstract}

\maketitle

\section{Introduction}
\label{sec_introduction}

Stimuli-responsive twist actuators upon activation show a net torsional deformation around one axis. Such devices could in the future, for instance, be used for the purpose of microfluidic mixing or during microsurgical operations \cite{aziz2019torsional}. To allow for a sufficient degree of softness of such objects and to facilitate the requested type of actuation, we have recently suggested to construct such actuators from magnetorheological gels or elastomers \cite{fischer2020towards}. These materials consist of magnetic or magnetizable colloidal particles enclosed by a permanently chemically crosslinked soft elastic polymeric environment \cite{filipcsei2007magnetic, jolly1996magnetoviscoelastic, ilg2013stimuli, 
li2014state, menzel2015tuned, odenbach2016microstructure, schmauch2017chained, weeber2018polymer,weeber2019studying, stolbov2019magnetostriction, menzel2019mesoscopic, schumann2019microscopic, zhou2019magnetoresponsive}. Upon magnetization by a homogeneous external magnetic field, such materials generally show an overall deformation of the magnetostrictive type. So far, mainly linear types of induced distortion have been considered in an external magnetic field, either homogeneous or inhomogeneous, implying extensions or contractions along straight global axes of deformation \cite{zrinyi1996deformation,
filipcsei2007magnetic, 
gollwitzer2008measuring, fuhrer2009crosslinking,ivaneyko2011magneto,stolbov2011modelling, 
maas2016experimental, metsch2016numerical, attaran2017modeling, fischer2019magnetostriction, gebhart2019general}.

To proceed one step further and to realize a magnetostrictive torsional response, we have suggested to consider globally twisted arrangements of the embedded magnetizable colloidal particles \cite{fischer2020towards}. When such systems are magnetized along the axis of twist, an overall torsional deformation is typically initiated by the particle structure in attempting to untwist itself. \am{This situation has been analyzed in detail in a study focusing on the mesoscopic scale, revealing, for instance, an optimum angle of initial twist to achieve the maximal induced torsional deformation\cite{fischer2020towards}. To this end, numerical evaluations for different discretized particle arrangements were performed.}

Different practical routes are conceivable to generate such systems of initially twisted particle structures. For example, the particles could be placed by hand or by a robot to the requested positions, while the surrounding polymeric environment is built up layer by layer \cite{chen2013numerical,puljiz2016forces,puljiz2018reversible}. A more refined strategy may combine methods that were developed in the two fields of magnetorheological elastomers and liquid-crystalline elastomers \cite{urayama2007selected,ohm2010liquid}. First, the magnetizable particles are dispersed in a suspension from which the polymeric elastic environment is generated through a subsequent chemical crosslinking reaction. If during this procedure a strong homogeneous magnetic field is applied, chain-like aggregates of the particles form that are then permanently maintained in the materials by the enclosing gel or elastomer \cite{collin2003frozen, coquelle2005magnetostriction,abramchuk2007novel,bose2007viscoelastic,chen2007microstructures,filipcsei2010magnetodeformation,gunther2011x,borbath2012xmuct, danas2012experiments}. After this first step of chemical crosslinking, the systems may be globally twisted by mechanical torsional forces applied from outside onto the samples. The corresponding axis of twist must be oriented parallel to the initial axes of the chain-like aggregates. If maintained in this twisted state while a second, retarded chemical crosslinking reaction is taking place, at least part of this overall torsional deformation can be locked into the materials. As a result, one obtains magnetorheological gels or elastomers containing globally twisted particle arrangements. Such a two-step crosslinking procedure has previously been applied successfully to generate monodomain nematic liquid-crystalline elastomers \cite{kupfer1991nematic,kupfer1994liquid,ohm2010liquid}, which feature outstanding mechanical properties \cite{kupfer1991nematic,kupfer1994liquid,urayama2007stretching, menzel2009response}. (Also the couplings between magnetic and deformational effects in combined magnetic and liquid crystalline systems were studied \cite{winkler2010liquid,siboni2020non, shrivastav2020steady}.) 

For not too large angles of initial twist and not too small initial distances between the particles, the magnetic interactions that are induced between the particles upon magnetization tend to revert the stored torsional deformation \cite{fischer2020towards}. To support a basic physical understanding and to allow for a comparison with possible corresponding experiments, we wish to quantify these induced deformations by appropriate mesoscopic and macroscopic theoretical approaches. Generally, magnetoelastic behavior is included in macroscopic continuum theories by a nonlinear magnetostrictive term that couples the macroscopic magnetization variable quadratically to the macroscopic strain tensor \cite{jarkova2003hydrodynamics, bohlius2004macroscopic}. Interestingly, in the present case, the induced twist deformation can be described in a reduced picture as well when these terms are neglected. 

To find an appropriate macroscopic description for twist-type deformations, including specifications of the macroscopic system parameters, we start by introducing an idealized discrete mesoscopic minimal model. We calculate within the framework of this model the magnetically induced distortions. On this basis, the necessary couplings to include the magnetostrictive torsional effects also into the macroscopic picture are illustrated. Moreover, within the context of the introduced assumptions, we find expressions of the macroscopic system parameters as functions of the mesoscopic model parameters. Consequently, an approximate scale bridging \cite{pessot2015towards, menzel2014bridging} is achieved. 

We start in Sec.~\ref{sec_macroproblem} by briefly pointing out the geometry of the considered system and the investigated types of elastic deformation. Afterwards, in Sec.~\ref{sec_mesomodel}, we introduce our discrete mesoscopic minimal model to describe the magnetically induced overall elastic deformations. On this basis, we turn in Sec.~\ref{sec_macroscopic} to the macroscopic characterization. Comparing the derived expressions on the two levels, mesoscopic and macroscopic, in Sec.~\ref{sec_scalebridging}, we reveal expressions for the associated macroscopic system parameters as functions of the mesoscopic model parameters. We can then outline in Sec.~\ref{sec_macrocomplete} a reduced but effective macroscopic picture to characterize the magnetically induced twist deformation in macroscopic experiments. Finally, we conclude in Sec.~\ref{sec_conclusions}.

\section{Cylindrical magnetoelastic example systems}
\label{sec_macroproblem}

\begin{figure}
\includegraphics[width=6.cm]{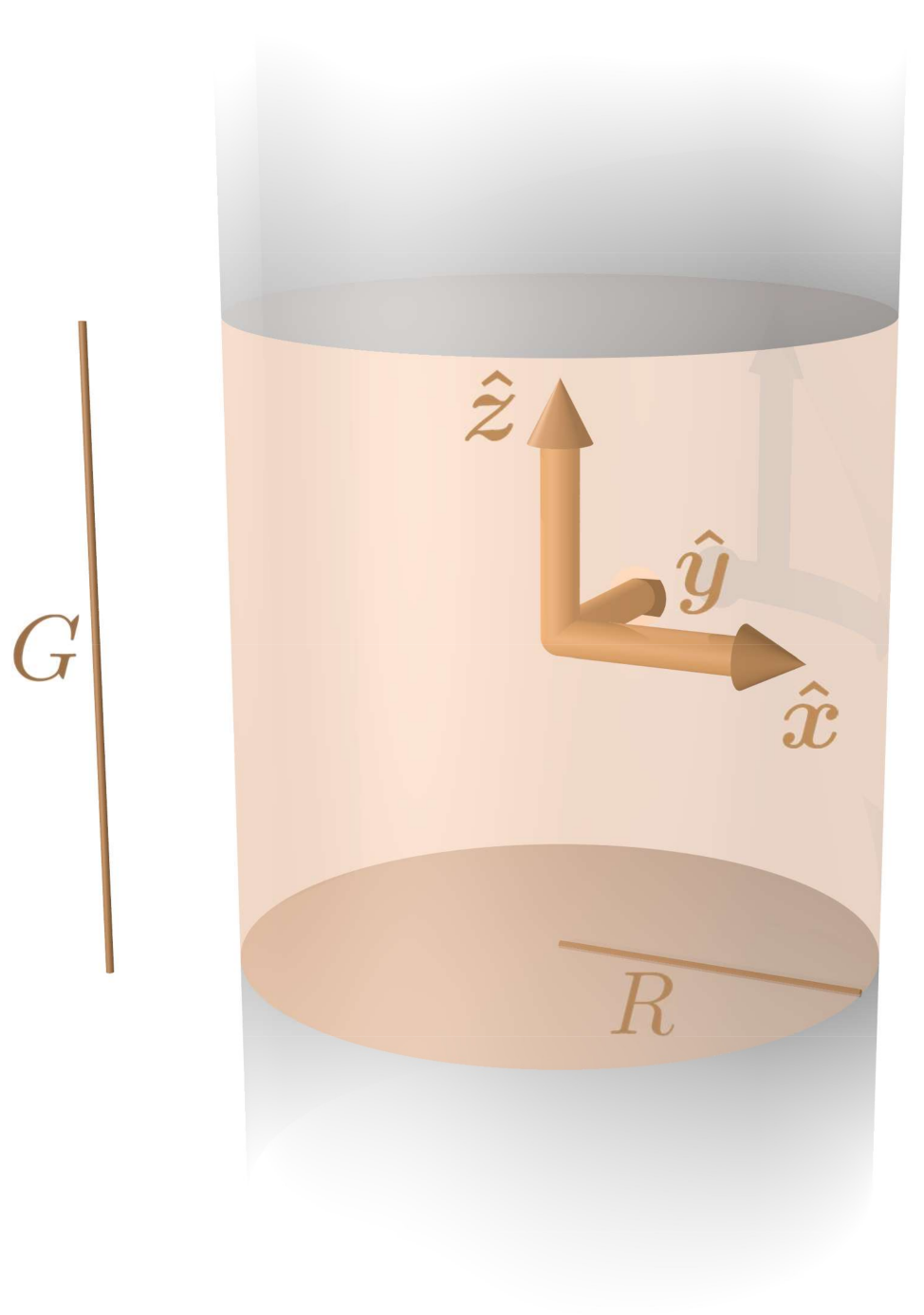}
\caption{We consider a system of cylindrical shape of height $G$, which is part of a cylinder of radius $R$. The cylindrical piece is centered around the origin of a Cartesian coordinate system, with the $z$-axis pointing along the vertically oriented cylindrical axis.}
\label{fig_cylinder_macro}
\end{figure}

We start by briefly introducing the types of magnetically induced elastic deformation that we focus on in this study. For illustration, we assume a piece of material of height $G$, which is part of a cylindrical sample of radius $R$, see Fig.~\ref{fig_cylinder_macro}. 
This geometry is in line with many practical realizations, where frequently samples of magnetorheological gels and elastomers are produced of cylindrical shape \cite{schumann2017situ,schumann2017characterisation, gundermann2017statistical}. We here consider three different types of macroscopic deformation of such samples: axial expansions or contractions along the cylinder axis, associated lateral contractions or expansions to the sides, and torsional (twist) deformations around the cylinder axis. Consequently, we parameterize the macroscopic strain tensor $\bm{\varepsilon}$ to quantify these types of distortion, see below. We assume the material to be incompressible, that is, the net volume of the cylindrical piece is preserved. Moreover, we confine ourselves to linearly elastic deformations of sufficiently small amplitudes \cite{landau1986theory}. 

Spatially inhomogeneous internal deformations, if present at all, are not resolved in this effective treatment. This point of view corresponds to an experimental characterization that only measures the overall axial expansions or contractions and net torsional deformations. Since the spatial details of the deformations are not resolved, we in effect treat the material as if it deformed in an affine way. 

Our coordinate system is chosen so that the $z$-axis coincides with the cylinder axis and the origin is located at the center of the cylindrical piece, see Fig.~\ref{fig_cylinder_macro}. The amplitude of contraction or expansion along the $z$-axis is denoted as $A$. We quantify the overall torsional deformation around the cylinder axis by $\tau=\Delta\Phi/G$, where $\Delta\Phi$ equals the angle of rotation of the top surface of the cylindrical piece relative to its bottom surface. For incompressible materials, the associated affine deformation can in Cartesian coordinates be characterized by a displacement field \cite{landau1986theory}
\begin{equation}\label{eq_u}
\mathbf{u}=\left(
\begin{array}{c}
-\frac{1}{2}Ax-\tau yz \\[.1cm]
-\frac{1}{2}Ay+\tau xz \\[.1cm]
Az
\end{array}
\right).
\end{equation}
In our analysis, we confine ourselves to small elastic deformations. Therefore, only the linearized elastic strain tensor is taken into account. Starting from the displacement field in Eq.~(\ref{eq_u}), we calculate the linearized strain tensor $\bm{\varepsilon}=\left(\bm{\nabla}\mathbf{u}+(\bm{\nabla}\mathbf{u})^T\right)/2$,\cite{landau1986theory} with $^T$ denoting the transpose, as 
\begin{equation}\label{eq_eps}
\bm{\varepsilon}=\left(
\begin{array}{ccc}
-\frac{1}{2}A & 0 & -\frac{1}{2}\tau y \\[.1cm]
0 & -\frac{1}{2}A & \frac{1}{2}\tau x \\[.1cm]
-\frac{1}{2}\tau y & \frac{1}{2}\tau x & A
\end{array}
\right).
\end{equation}
From here, integrating the purely elastic part of the associated free-energy density \cite{landau1986theory} over the whole cylinder, we obtain the resulting elastic free energy of deformation as 
\begin{equation}\label{eq_F_el}
{\mathcal{F}_{\mathrm{el}}}
=\int\mathrm{d}^3\!r\,\mu\,\bm{\varepsilon}:\bm{\varepsilon}
=\frac{3}{2}\pi\mu\, GR^2A^2 + \frac{1}{4}\pi\mu\, G\tau^2 R^4,
\end{equation}
where $\mu$ sets the effective elastic shear modulus of the elastic material. 


Concerning its magnetic properties, we consider the system to be genuinely paramagnetic (or superparamagnetic) without any or only negligible macroscopic remnant magnetization. An external magnetic field $\mathbf{H}=H\mathbf{\hat{z}}$ is applied to magnetize the system. We assume the external magnetic field to remain weak enough so that we stay within the window of linear response of the induced magnetization $\mathbf{M}$ with respect to $\mathbf{H}$. 

To lowest order, we might further assume that $\mathbf{M}\parallel\mathbf{H}\parallel\mathbf{\hat{z}}$ in our geometry. Generally, magnetostrictive contributions to the macroscopic free-energy density explicitly couple the strain tensor and the magnetization to each other \cite{jarkova2003hydrodynamics,bohlius2004macroscopic}. For our system and for $\mathbf{M}\parallel\mathbf{\hat{z}}$, the corresponding overall expression \cite{bohlius2004macroscopic} then could be rewritten as 
\begin{equation}\label{eq_Fmstrav}
\tilde{F}_{\mathrm{mstr}} =
{}-\frac{1}{2}\,\tilde{\zeta}\,\mathbf{M}\cdot
\bm{\varepsilon}\cdot\mathbf{M}
= {}-\frac{1}{2}\,\tilde{\zeta}\, A\,M^2,
\end{equation}
with $\tilde{\zeta}$ denoting the corresponding effective system parameter. 
Yet, in the present context, such a framework seems to be incomplete. Depending on the sign of $\tilde{\zeta}$, only linear expansions or contractions of amplitude $A$ along the direction of the magnetic field are described (together with associated lateral contractions or expansions, respectively). There is no coupling to the torsional amplitude $\tau$. Consequently, magnetostrictive twist actuation is not covered by this simplified level of representation. How do we have to formulate our characterization to include this effect?

\section{Discrete mesoscopic minimal model including magnetically induced torsional deformations}
\label{sec_mesomodel}

To find an answer to the question raised above, we here introduce a basic mesoscopic minimal model. Our approach is discrete in the sense that we consider a set of identical spherical inclusions (particles) in the continuous elastic environment. For simplicity, the separation between the inclusions is large enough so that they can be treated as point-like. They are perfectly paramagnetic (or superparamagnetic) and do not show any remnant magnetization. Weak external magnetic fields allow us to use linear magnetization laws, and we denote the magnetic bulk susceptibility of the particle material as $\chi$. Moreover, we assume the elastic polymer body to only deform affinely, which represents an approximation at this level \cite{pessot2014structural} for the specified particle arrangement (see below) and renders our analysis semi-quantitative. \am{The presence of boundaries can further promote inhomogeneous deformations under many-body magnetic interactions \cite{goh2018dynamics}.}

\begin{figure*}
\includegraphics[width=15.cm]{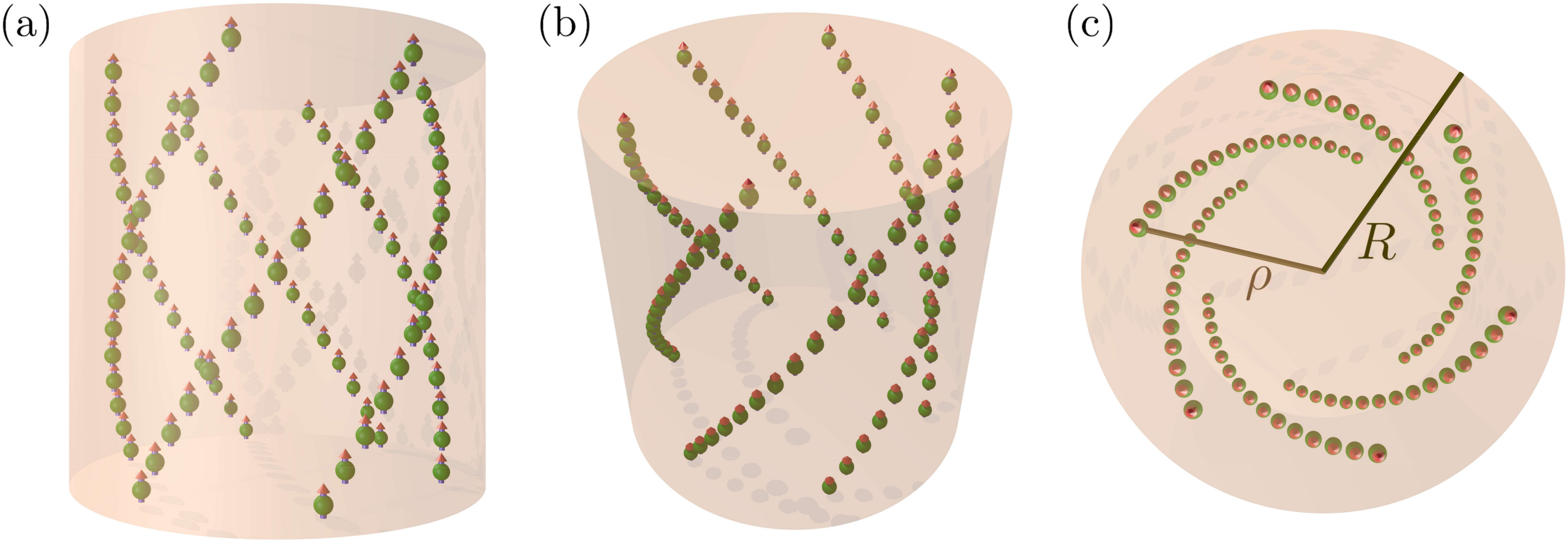}
\caption{Illustration of the minimal model picture introduced to achieve our scale connection. A soft elastic polymeric body of cylindrical shape contains discrete chain-like particle aggregates that are twisted around the cylindrical axis. In our calculations, we concentrate on a cylindrical piece of height $G$ of a much more elongated cylinder, see Fig.~\ref{fig_cylinder_macro}. Therefore, boundary effects on the top and bottom surfaces are neglected. The radius of the cylindrical elastic body is denoted as $R$, while all particles take the same radial distance $\rho$ from the cylinder axis. Induced magnetic moments of the particles are indicated by a small arrow on each inclusion. (a) Side view, (b) top-side view, and (c) top view.}
\label{fig_geometry}
\end{figure*}

\begin{figure}
\includegraphics[width=6.cm]{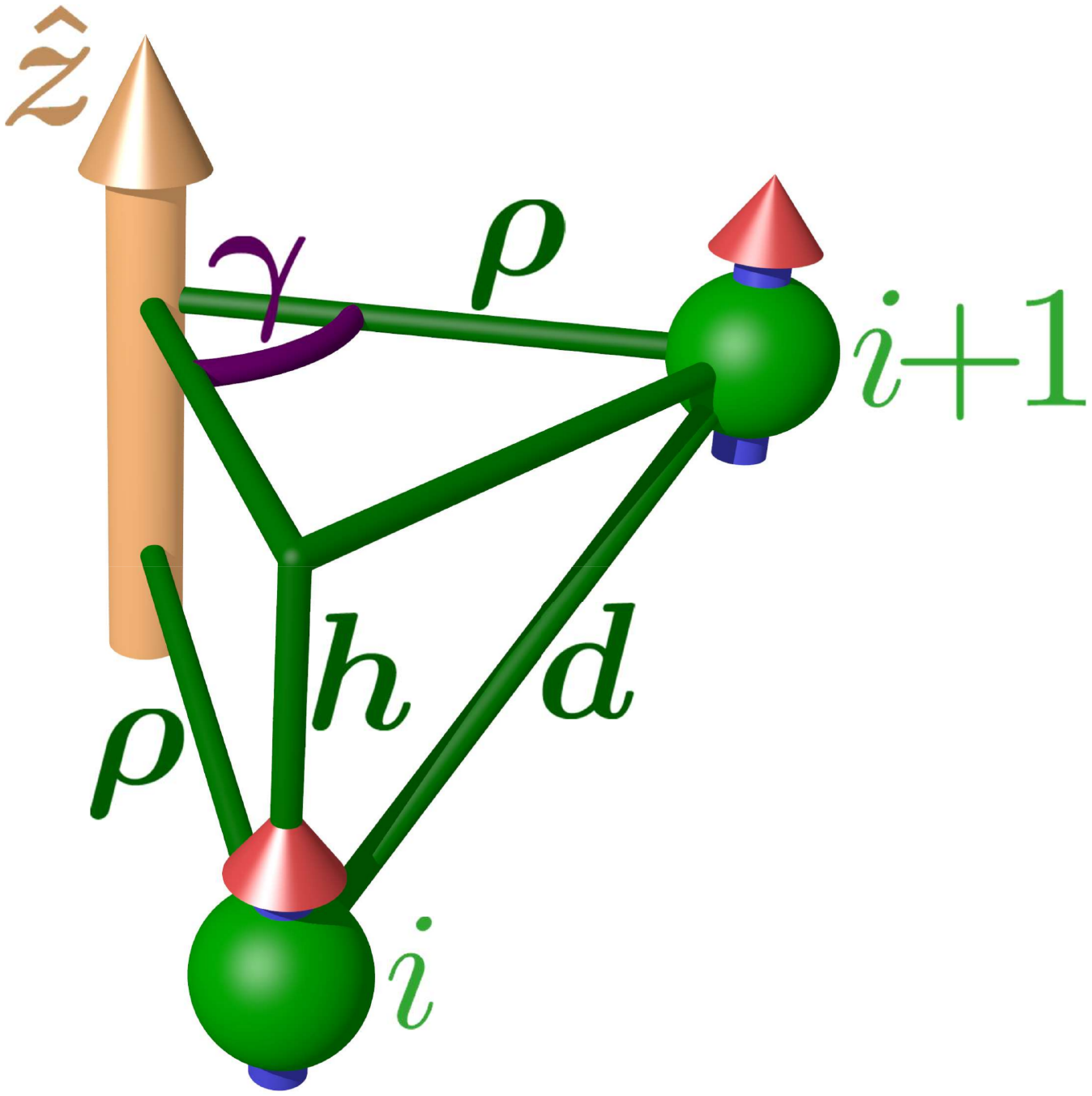}
\caption{Definition of the quantities used to parameterize the twisted mesoscopic particle structures. The twist axis coincides with the $\mathbf{\hat{z}}$ direction, along which also the external magnetic field $\mathbf{H}$ is applied. All particles take a distance $\rho$ from the twist axis. From one particle $i$ to the next particle $i+1$ above it on the same twisted chain-like particle aggregate, the vertical distance along $\mathbf{\hat{z}}$ is denoted as $h$, while the actual distance between the two particles is referred to as $d$. The twist angle between the radial position vectors of any such two vertically neighboring particles is called $\gamma$.}
\label{fig_cylinder_meso}
\end{figure}

In principle, we here consider a piece of height $G$ being part of a cylinder that is infinitely extended along its center axis. That is, we concentrate on the inner part of the sample and neglect boundary effects at the top and bottom surfaces. All particles are organized in chain-like aggregates and \am{for simplicity are assumed to} take the same distance $\rho$ from the cylinder axis, see Fig.~\ref{fig_geometry}. 
In the initial nonmagnetized state of the system, the particle chains are homogeneously twisted around the cylinder axis. The twist angle $\gamma$ quantifies how the radial position vector of the particles is rotated around the cylinder axis from one particle to the next upper one. In the vertical direction, all particles are separated by the same vertical distance $h$, see Fig.~\ref{fig_cylinder_meso}. 
It follows that the distance $d$ between two adjacent particles is given by 
\begin{equation}
d=\sqrt{h^2+4\rho^2\sin^2\!\frac{\gamma}{2}}\,.
\end{equation}

There are two types of contribution to the magnetization of each particle. First, the external magnetic field $\mathbf{H}$ leads to a direct magnetization and a resulting magnetic dipole moment 
\begin{equation}\label{eq_m0}
\mathbf{m}_0=\frac{4\pi}{3}a^3
\frac{3\chi}{\chi+3}\mathbf{H}.
\end{equation}
Here, $a$ denotes the radius of the spherical inclusions. The additional factor of $3/(\chi+3)$ is associated with the demagnetization effect in a spherical particle \cite{jackson1962classical, puljiz2018reversible}. 

The second type of contribution to the magnetic moment arises from the mutual magnetization between the magnetized particles. This is a nonlinear effect in the sense that it is at least quadratic in $\chi$. Here, we only consider the lowest-order contribution arising from the additional magnetization by the induced magnetic moments $\mathbf{m}_0$ of the nearest-neighboring particles on the same chain. We obtain the corresponding additional contribution $\mathbf{m}_1$ to the magnetic moment in the same way as $\mathbf{m}_0$ in Eq.~(\ref{eq_m0}). However, we there have to insert the magnetic field induced by the nearest-neighboring magnetic dipole moments $\mathbf{m}_0$ instead of the external magnetic field $\mathbf{H}$ on the right-hand side. Along these lines, we find
\begin{equation}\label{eq_m1}
\mathbf{m}_1 = \frac{2\chi}{\chi+3}\frac{a^3}{d^5}m_0
\left[3h\rho\sin\!\gamma\,\bm{\hat{\varphi}}
+2\left(h^2-2\rho^2\sin^2\!\frac{\gamma}{2}\right)\mathbf{\hat{z}} \right], 
\end{equation}
where $m_0=\|\mathbf{m}_0\|$ \am{denotes the magnitude of $\mathbf{m}_0$. Moreover,} $\bm{\hat{\varphi}}$ is the polar unit vector perpendicular to $\mathbf{\hat{z}}$ and tangential to the enveloping cylinder of the twisted helical particle structure. This tangential component arises directly from the mutual tangential shift into the polar direction between the particles separated by a vertical distance $h$. Conversely, the vertical component of $\mathbf{m}_1$ along $\mathbf{\hat{z}}$ is the obvious one resulting from the vertical separation between the nearest-neighboring inclusions. There is no radial contribution because the outward component induced by the magnetic field of the lower nearest-neighboring magnetized inclusion is balanced by the inward component induced by the upper nearest neighbor. 

For each particle, the change in magnetic energy when bringing the sample into the homogeneous external magnetic field $\mathbf{H}$ to the considered order reads \cite{puljiz2018reversible}
\begin{equation}\label{eq_WmB}
\mathcal{W}_{\mathrm{mag}}^{\,\mathrm{particle}} = 
{}-\frac{\mu_0}{2}(\mathbf{m}_0+\mathbf{m}_1)\cdot\mathbf{H}, 
\end{equation}
where $\mu_0$ denotes the magnetic vacuum permeability. 
From the number of particles per chain within the cylindrical piece of height $G$ and the number of chains within this cylinder we can find the total number $N$ of particles. Thus, the particle concentration is given by $n=N/\pi R^2G$. Adding the contributions from affinely deforming the elastic material, see Eq.~(\ref{eq_F_el}), the overall energy per volume is therefore obtained as 
\begin{equation}\label{eq_Wprev}
\frac{\mathcal{W}}{\pi R^2G} =
\frac{3}{2}\mu A^2 + \frac{1}{4}\mu\tau^2R^2
-\frac{\mu_0}{2}n(\mathbf{m}_0+\mathbf{m}_1)\cdot\mathbf{H}.
\end{equation}

We now denote in the initial, nonmagnetized, undeformed state of the material the vertical distance between the particles, their radial distance from the cylinder axis, and the polar angle between their neighboring position vectors as $h_0$, $\rho_0$, and $\gamma_0$, respectively. Consequently, we can define the pitch of the helix in the undistorted state via 
\begin{equation}\label{eq_q0}
q_0=\frac{\gamma_0}{h_0}.
\end{equation}
From there, in the distorted state, we obtain the corresponding quantities 
\begin{eqnarray}
h &=& h_0\left(1+A\right), \label{eq_h}\\[.1cm]
\rho &=& \rho_0\left(1-\frac{A}{2}\right), \label{eq_rho} \\[.1cm]
\gamma &=& \left(q_0+\tau\right) h_0 
\label{eq_gamma}
\end{eqnarray}
up to linear order in the amplitudes of deformation, see also Appendix~\ref{app_quantities}. 
When inserting these expressions into Eq.~(\ref{eq_Wprev}), we find
\begin{widetext}
\begin{equation}
\frac{\mathcal{W}}{\pi R^2G} =
\frac{3}{2}\mu A^2 + \frac{1}{4}\mu\tau^2R^2 
-2\pi na^3\frac{\chi}{\chi+3}\mu_0{H^2}
\Bigg[1+4a^3\frac{\chi}{\chi+3}\frac{h_0^2(1+A)^2-2\rho_0^2\left(1-\frac{A}{2}\right)^2
\sin^2\!\frac{(q_0+\tau)h_0}{2}}{
\sqrt{h_0^2(1+A)^2+4\rho_0^2\left(1-\frac{A}{2}\right)^2
\sin^2\!\frac{(q_0+\tau)h_0}{2}}^5}\Bigg]. 
\label{eq_W_befexp}
\end{equation}
\am{If additional chain-like aggregates of different distance from the cylinder axis are considered in the mesoscopic model, their energetic contributions are included by analogous terms.} 

To proceed further, we expand the last term in Eq.~(\ref{eq_W_befexp}) up to linear order in $A$ and $\tau$, leading to
\begin{eqnarray}\label{eq_meso_exp}
\frac{\mathcal{W}}{\pi R^2G} &=&
\frac{3}{2}\mu A^2 + \frac{1}{4}\mu\tau^2R^2 
-2\pi na^3\frac{\chi}{\chi+3}\mu_0{H^2}
\Bigg[1+4a^3\frac{\chi}{\chi+3}\frac{h_0^2-2\rho_0^2
\sin^2\!\frac{q_0h_0}{2}}{
\sqrt{h_0^2+4\rho_0^2\sin^2\!\frac{q_0h_0}{2}}^5}\Bigg]
\nonumber\\
&&{}
+24\pi na^6\left(\frac{\chi}{\chi+3}\right)^{\!2}\mu_0{H^2}
\frac{ 4\rho_0^4\sin^4\!\frac{q_0h_0}{2}-10h_0^2\rho_0^2\sin^2\!\frac{q_0h_0}{2}
\!+\!h_0^4 
}{
\sqrt{h_0^2+4\rho_0^2
\sin^2\!\frac{q_0h_0}{2}}^7}\,A
\nonumber\\
&&{}
+48\pi na^6\left(\frac{\chi}{\chi+3}\right)^{\!2}\mu_0{H^2}\,
h_0\rho_0^2\frac{\left(h_0^2-\rho_0^2\sin^2\!\frac{q_0h_0}{2}\right)\sin(q_0h_0)}{\sqrt{h_0^2+4\rho_0^2\sin^2\!\frac{q_0h_0}{2}}^7}\,\tau.
\end{eqnarray}
From here, we find the amplitude $A$ of contractile and extensile deformations via $\partial\mathcal{W}/\partial A=0$ as
\begin{equation}
A = {}-8\pi na^6\left(\frac{\chi}{\chi+3}\right)^{\!2}
\frac{ 4\rho_0^4\sin^4\!\frac{q_0h_0}{2}-10h_0^2\rho_0^2\sin^2
\!\frac{q_0h_0}{2} 
\!+\!h_0^4 
}{\mu
\sqrt{h_0^2+4\rho_0^2
\sin^2\!\frac{q_0h_0}{2}}^7}\,\mu_0{H^2},
\end{equation}
while the amplitude $\tau$ of twist deformation follows via $\partial\mathcal{W}/\partial\tau=0$ as
\begin{equation}
\tau={}-96\pi na^6\left(\frac{\chi}{\chi+3}\right)^{\!2}\,
h_0\rho_0^2\frac{\left(h_0^2-\rho_0^2\sin^2\!\frac{q_0h_0}{2}\right)\sin(q_0h_0)}{\mu R^2\sqrt{h_0^2+4\rho_0^2\sin^2\!\frac{q_0h_0}{2}}^7}\,\mu_0{H^2}.
\end{equation}
Obviously, both $A\propto H^2$ and $\tau\propto H^2$. Moreover, without any initial twist of the internal structure, that is, for $q_0=0$, we do not find any magnetically induced torsional deformation, implying $\tau=0$. In contrast to that, still a magnetoelastic deformation $A\neq0$ is observed for $q_0=0$ and $H\neq0$. 

For completeness and for later comparison, we list the components of the magnetization $\bm{\mathcal{M}}^{\mathrm{meso}}=n(\mathbf{m}_0+\mathbf{m}_1)$ resulting from our mesoscopic model. Specifically, from Eqs.~(\ref{eq_m0}), (\ref{eq_m1}), and (\ref{eq_h})--(\ref{eq_gamma}) we find
\begin{eqnarray}
\mathcal{M}^{\mathrm{meso}}_{\varphi} &=&
24\pi na^6\left(\frac{\chi}{\chi+3}\right)^2
\frac{h_0\rho_0\sin(q_0h_0)}{\sqrt{h_0^2+4\rho_0^2\sin^2\!\frac{q_0h_0}{2}}^5}\,H
\nonumber\\[.1cm]
&&{}
+36\pi na^6\left(\frac{\chi}{\chi+3}\right)^2h_0\rho_0
\frac{8\rho_0^2\sin^2\!\frac{q_0h_0}{2}-3h_0^2}{\sqrt{h_0^2+4\rho_0^2\sin^2\!\frac{q_0h_0}{2}}^7}\sin\left(q_0h_0\right)\,H\,A
\nonumber\\
&&{}
+12\pi na^6\left(\frac{\chi}{\chi+3}\right)^2h_0^2\rho_0
\frac{2h_0^2\cos(q_0h_0)-\rho_0^2\left(9\sin\frac{q_0h_0}{2}+\sin\frac{3q_0h_0}{2}\right)}{\sqrt{h_0^2+4\rho_0^2\sin^2\!\frac{q_0h_0}{2}}^7}\,H\,\tau.
\label{eq_Mmeso_phi}
\end{eqnarray}
Moreover, combining the definition of $\bm{\mathcal{M}}^{\mathrm{meso}}$ with Eq.~(\ref{eq_Wprev}) and taking into account $\mathbf{H}=H\mathbf{\hat{z}}$, we obtain 
\begin{equation}\label{eq_Mmeso_z}
\mathcal{M}^{\mathrm{meso}}_z=
{}-\frac{2}{\mu_0H} \left( \frac{\mathcal{W}}{\pi R^2G} -\frac{3}{2}\mu A^2 - \frac{1}{4}\mu\tau^2R^2 \right),
\end{equation}
where an explicit expression follows by inserting  Eq.~(\ref{eq_meso_exp}). 
\end{widetext}

\section{Macroscopic description including magnetoelastic twist deformations}
\label{sec_macroscopic}

We now start from our insight on the level of the mesoscopic picture and the expression for the energy per cylindrical volume in Eq.~(\ref{eq_meso_exp}) to develop an appropriate macroscopic description. In this way, we can later identify expressions of the macroscopic system parameters in terms of the mesoscopic model parameters. Accordingly, in Sec.~\ref{sec_scalebridging}, this step will link the two different scales of description to each other. 

First, we observe that in the mesoscopic model the contribution $\mathbf{m}_1$ in Eq.~(\ref{eq_m1}) to the magnetic moment resulting from the mutual particle magnetization plays an essential role. It is this contribution that involves the twist angle $\gamma$. The angle $\gamma$ is related to the oblique orientation of $\mathbf{m}_1$ relative to the vertical axis $\mathbf{\hat{z}}$. To make progress, we therefore must search to incorporate this effect into the macroscopic characterization. 

For this purpose, we take into account the local anisotropy axis $\mathbf{\hat{n}}$ associated with the local orientation of the twisted particle structures. In a corresponding mesoscopic picture, $\mathbf{\hat{n}}$ is assumed to be locally tangential to the perfect twisted chain-like particle aggregates. Therefore, we may parameterize it macroscopically as
\begin{equation}\label{eq_n}
\mathbf{\hat{n}}=n_{\varphi}\bm{\hat{\varphi}}+n_z\bm{\hat{z}}.
\end{equation}

Obviously, the tilt of $\mathbf{\hat{n}}$ into the direction $\bm{\hat{\varphi}}$ is of central importance. In the macroscopic picture, it breaks the vertical mirror symmetry of the cylinder. It shall therefore be taken into account to include the twist actuation. Particularly, it will describe the emergence of a component of the macroscopic magnetization along $\bm{\hat{\varphi}}$. For an effective characterization, we turn to cylindrical coordinates. 

In contrast to the mesoscopic model in Sec.~\ref{sec_mesomodel}, where we assumed all magnetic particles to feature one identical distance $\rho$ from the cylinder axis, we now consider the magnetizable structural elements to be distributed over the volume of the cylinder in the macroscopic approach. Again, the internal structures show an intrinsic initial global twist in the undeformed state. Simultaneously, deformations of the cylinder can also act to reorient $\mathbf{\hat{n}}$. 

The tilt $n_{\varphi}$ of $\mathbf{\hat{n}}$ relative to $\mathbf{\hat{z}}$, see Eq.~(\ref{eq_n}), is a function of the location within the cylinder. In cylindrical coordinates, it only depends on the radial distance $r$ from the cylinder axis. To calculate $\mathbf{\hat{n}}(r)$, we make use of Eqs.~(\ref{eq_h})--(\ref{eq_gamma}) already derived in Sec.~\ref{sec_mesomodel}. We consider a vertical vector $\mathbf{\tilde{n}}=\mathbf{r}_{\mathrm{end}}-\mathbf{r}_{\mathrm{start}}$ from position 
\begin{equation}
\mathbf{r}_{\mathrm{start}}=\left(\begin{array}{c}r\\[.1cm]0\\[.1cm]0\end{array}\right)
\end{equation}
to position 
\begin{equation}
\mathbf{r}_{\mathrm{end}}=\left(\begin{array}{c}r\\[.1cm]0\\[.1cm]\tilde{h}\end{array}\right), 
\end{equation}
defined even before the initial global twist of the structure is implemented. Next, the initial global twist and some additional elastic deformations, quantified by $A$ and $\tau$, are imposed. Thus, the two positions $\mathbf{r}_{\mathrm{start}}$ and $\mathbf{r}_{\mathrm{end}}$ are shifted to 
\begin{equation}
\mathbf{r}_{\mathrm{start}}'=\left(\begin{array}{c} 
r\left(1-\frac{A}{2}\right) \\[.1cm] 0 \\[.1cm]0 \end{array}\right)
\end{equation}
and
\begin{equation}
\mathbf{r}_{\mathrm{end}}'=\left(\begin{array}{c}
r\left(1-\frac{A}{2}\right) \\[.1cm]
r\left(1-\frac{A}{2}\right)(q_0+\tau)\tilde{h} \\[.1cm] 
\tilde{h}(1+A)
\end{array}\right), 
\end{equation} 
respectively, where we restrict the magnitude of $\tilde{h}$ so that $(q_0+\tau)\tilde{h}\ll1$ is implied. 
From here, we calculate $\mathbf{\tilde{n}}'=\mathbf{r}_{\mathrm{end}}'-\mathbf{r}_{\mathrm{start}}'$. After normalization, the components of $\mathbf{\hat{n}}(r)$ are identified as 
\begin{eqnarray}
\label{eq_n_phi}
\hspace{-.6cm}n_{\varphi}&=&\mathrm{sgn}(q_0)\sqrt{1-n_z^2}, \\
\label{eq_n_z}
\hspace{-.6cm}n_z &=& \frac{1}{\sqrt{1\!+\!q_0^2r^2}}
+\frac{3q_0^2r^2}{2\sqrt{1\!+\!q_0^2r^2}^3}\,A
-\frac{q_0r^2}{\sqrt{1\!+\!q_0^2r^2}^3}\,\tau,
\end{eqnarray}
where $\mathrm{sgn}$ denotes the sign function and here returns the sign of $q_0$. 

In analogy to Eq.~(\ref{eq_eps}), the strain tensor in cylindrical coordinates reads
\begin{equation}\label{eq_eps_cyl}
\bm{\varepsilon}=\left(
\begin{array}{ccc}
-\frac{1}{2}A & 0 & 0 \\[.1cm]
0 & -\frac{1}{2}A & \frac{1}{2}\tau r \\[.1cm]
0 & \frac{1}{2}\tau r & A
\end{array}
\right),
\end{equation}
with the first, second, and third row or column being associated with the components along $\mathbf{\hat{r}}$, $\bm{\hat{\varphi}}$, and $\mathbf{\hat{z}}$, respectively. Similarly, we obtain for the magnetization
\begin{equation}\label{eq_M_cly}
\mathbf{M}=M_{\varphi}\bm{\hat{\varphi}}+M_z\mathbf{\hat{z}}.
\end{equation}

As before, we do not take into account any anisotropy of the purely elastic contributions to the macroscopic free-energy density. For small enough volume fractions and a significant separation between the magnetizable particles as in our mesoscopic model picture, see Fig.~\ref{fig_cylinder_meso}, this approximation appears justified. Thus, using $\bm{\varepsilon}$ from Eq.~(\ref{eq_eps_cyl}), we obtain the elastic part of the free-energy density of the form 
\begin{equation}\label{eq_Fel}
F_{\mathrm{el}}=\mu\,\bm{\varepsilon}:\bm{\varepsilon}
=\frac{3}{2}\mu A^2 + \frac{1}{2}\mu r^2\tau^2.
\end{equation}
From here, we find the same result for the purely elastic part $\mathcal{F}_{\mathrm{el}}$ of the macroscopic free energy as in Eq.~(\ref{eq_F_el}). 

Next, we turn to the purely magnetic part of the macroscopic free-energy density
, which reads
\begin{equation}\label{eq_Fm}
F_{\mathrm{M}}={}\frac{1}{2}\bm{\alpha}:\mathbf{M}\mathbf{M}
-\mu_0\,\mathbf{M}\cdot\mathbf{H}.
\end{equation}
The structural anisotropy enters this expression via 
\begin{equation}\label{eq_alpha}
\bm{\alpha}=\alpha_{\|}\mathbf{\hat{n}}\mathbf{\hat{n}}
+\alpha_{\bot}\!\left(\mathbf{I}-\mathbf{\hat{n}}\mathbf{\hat{n}}\right)
\end{equation}
with the two scalar material coefficients $\alpha_{\|}$ and $\alpha_{\bot}$ as well as the identity matrix $\mathbf{I}$. 
For a genuinely paramagnetic system, it follows that $\alpha_{\|}>0$ and $\alpha_{\bot}>0$. 
The coupling term $-\,\mu_0\,\mathbf{M}\cdot\mathbf{H}$ expresses the influence of the external magnetic field $\mathbf{H}$ onto the system, see also Appendix~\ref{app_a}, by introducing a nonvanishing magnetization $\mathbf{M}$. We remark that frequently an additional term $\frac{1}{4}\beta\left((\mathbf{M})^2\right)^{\!2}$ is added to the magnetic part of the free-energy density \cite{jarkova2003hydrodynamics}. A large enough $\beta>0$ guarantees thermodynamic stability in combination with the nonlinear magnetostrictive term introduced below. Particularly, a spontaneously arising large magnetization coupled to a spontaneously arising large deformation, both of which not induced or enforced from outside, is bounded by this quartic contribution. 
Here, we consider the regime of small magnitudes of magnetization that are linearly generated by the external magnetic field from the nonmagnetized initial state. Therefore, the magnitudes of deformation and magnetization are assumed to be small enough so that the quadratic terms in Eqs.~(\ref{eq_Fel}) and (\ref{eq_Fm}) stabilize the system around the initial state \cite{bohlius2004macroscopic}. Thus, the quartic term in $\mathbf{M}$ is not taken into account. 

Finally, the mentioned magnetostrictive term in the free-energy density reads
\begin{equation}\label{eq_Fmstr}
F_{\mathrm{mstr}}={}-\frac{1}{2}\,\zeta_{ijkl}\,M_i\varepsilon_{jk}M_l, 
\end{equation}
where we use Einstein's summation convention. This term explicitly couples the strain tensor to the magnetization \cite{jarkova2003hydrodynamics, bohlius2004macroscopic}. In our locally uniaxial system, we find for the components of the fourth-rank magnetostrictive material tensor \cite{bohlius2004macroscopic}
\begin{eqnarray}\label{eq_zeta}
\zeta_{ijkl}&=&\zeta_1\,n_in_jn_kn_l + \zeta_2\,n_in_l\delta_{jk} + \zeta_3\,\delta_{il}n_jn_k
\nonumber\\&&{}
+\frac{1}{2}\zeta_4\,(n_in_j\delta_{kl}+n_in_k\delta_{jl}+n_jn_l\delta_{ik}+n_kn_l\delta_{ij}) 
\nonumber\\&&{}
+\zeta_5\,\delta_{il}\delta_{jk}
+\frac{1}{2}\zeta_6\,(\delta_{ij}\delta_{kl}+\delta_{ik}\delta_{jl}),
\end{eqnarray}
with $\delta_{ij}$ denoting the Kronecker delta. Out of these contributions, the terms $\propto\!\zeta_2$ and $\propto\!\zeta_5$ are dropped in the following because they would include the vanishing trace of $\bm{\varepsilon}$. 

Summing up the contributions according to Eqs.~(\ref{eq_Fel}), (\ref{eq_Fm}), and (\ref{eq_Fmstr}), the overall macroscopic free-energy density becomes
\begin{eqnarray}\label{eq_F_macro}
F^{\,\mathrm{macro}} &=& \frac{3}{2}\,\mu\,A^2+\frac{1}{2}\,\mu\,r^2\tau^2 
+\frac{1}{2}\,\bm{\alpha}:\mathbf{M}\mathbf{M}
\nonumber\\[.1cm]
&&{}
-\frac{1}{2}\,\bm{\zeta}\;\fourvdots\;\mathbf{M}\bm{\varepsilon}\mathbf{M}
-\mu_0\,\mathbf{M}\cdot\mathbf{H}.
\end{eqnarray}
From here, the free energy is obtained as $\mathcal{F}^{\,\mathrm{macro}}=\int \mathrm{d}^3r\,F^{\,\mathrm{macro}}$. Dividing $\mathcal{F}^{\,\mathrm{macro}}$ by the volume of the cylinder, we find from the first two, purely elastic contributions on the right-hand side of Eq.~(\ref{eq_F_macro}) the same two terms as on the right-hand side of Eq.~(\ref{eq_meso_exp}). 

An expression for $\mathbf{M}$ as a function of $\mathbf{H}$ can be derived by minimizing 
\begin{equation}
\frac{\delta \mathcal{F}^{\,\mathrm{macro}}}{\delta \mathbf{M}} = \mathbf{0},
\end{equation}
which here is identical to requiring
\begin{equation}\label{eq_M_implicit}
\frac{\partial F^{\,\mathrm{macro}}}{\partial \mathbf{M}} = \mathbf{0}.
\end{equation}
This implies 
\begin{equation}\label{eq_M_solvefor}
\alpha_{\bot}\big(\mathbf{A}-\bm{\Gamma}(\bm{\varepsilon})\big)\cdot\mathbf{M}=\mu_0\mathbf{H},
\end{equation}
\\
where we have defined
\begin{eqnarray}
\mathbf{A} &=& \mathbf{I}-\mathbf{\hat{n}}\mathbf{\hat{n}}+\frac{\alpha_{\|}}{\alpha_{\bot}}\mathbf{\hat{n}}\mathbf{\hat{n}},
\\
\bm{\Gamma}(\bm{\varepsilon}) &=&
\frac{\zeta_1}{\alpha_{\bot}}\mathbf{\hat{n}}\cdot\bm{\varepsilon}\cdot\mathbf{\hat{n}}\,\mathbf{\hat{n}}\mathbf{\hat{n}}
+\frac{\zeta_3}{\alpha_{\bot}}\mathbf{\hat{n}}\cdot\bm{\varepsilon}\cdot\mathbf{\hat{n}}\,\mathbf{I} 
\nonumber\\
&&{}
+\frac{\zeta_4}{\alpha_{\bot}}\mathbf{\hat{n}}\mathbf{\hat{n}}\cdot\bm{\varepsilon}+\frac{\zeta_4}{\alpha_{\bot}}(\mathbf{\hat{n}}\cdot\bm{\varepsilon})\mathbf{\hat{n}}
+\frac{\zeta_6}{\alpha_{\bot}}\bm{\varepsilon}.
\end{eqnarray}
Solving Eq.~(\ref{eq_M_solvefor}) for $\mathbf{M}$, we find
\begin{equation}\label{eq_sol_M}
\mathbf{M}=\frac{1}{\alpha_{\bot}}\sum_{p=0}^{\infty}\left[\mathbf{A}^{-1}\cdot\bm{\Gamma}(\bm{\varepsilon})\right]^p\cdot
\mathbf{A}^{-1}\cdot\mu_0\mathbf{H},
\end{equation}
where 
\begin{equation}
\mathbf{A}^{-1}=\mathbf{I}+\frac{\alpha_{\bot}-\alpha_{\|}}{\alpha_{\|}}\mathbf{\hat{n}}\mathbf{\hat{n}}.
\end{equation}

In the following, we confine ourselves to the lowest and linear order of $\mathbf{M}$ in $A$ and $\tau$, in agreement with our considerations of $\bm{\mathcal{M}}^{\mathrm{meso}}$ in Sec.~\ref{sec_mesomodel}, see Eqs.~(\ref{eq_Mmeso_phi}) and (\ref{eq_Mmeso_z}). 

\begin{widetext}
Inserting the expressions for the components of $\mathbf{\hat{n}}$, see Eqs.~(\ref{eq_n_phi}) and (\ref{eq_n_z}), as well as for $\mathbf{M}$, see Eq.~(\ref{eq_sol_M}), together with Eqs.~(\ref{eq_alpha}) and (\ref{eq_zeta}) into Eq.~(\ref{eq_F_macro}), we find for the macroscopic free-energy density up to the same order as for the mesoscopic energy density in Eq.~(\ref{eq_meso_exp})
\begin{eqnarray}
F^{\,\mathrm{macro}} &=& \frac{3}{2}\,\mu\,A^2+\frac{1}{2}\,\mu\,r^2\tau^2 
+\frac{1}{2}\left\{-\,\frac{1}{\alpha_{\bot}}
+\frac{\alpha_{\|}-\alpha_{\bot}}{\alpha_{\|}\alpha_{\bot}}
\frac{1}{1+q_0^2r^2}\right\}\mu_0^2H^2
\hspace{5.5cm}
\nonumber
\end{eqnarray}
\begin{eqnarray}
\qquad\qquad\qquad
&&{}
+\frac{1}{2}\bigg\{
\frac{\alpha_{\|}-\alpha_{\bot}}{\alpha_{\|}\alpha_{\bot}}
\frac{3q_0^2r^2}{\left(1+q_0^2r^2\right)^2}
-\frac{\zeta_1\alpha_{\bot}^2+\zeta_3(\alpha_{\bot}^2-\alpha_{\|}^2)+2\zeta_4\alpha_{\bot}({\alpha_{\bot}}-\alpha_{\|})+\zeta_6(\alpha_{\bot}-\alpha_{\|})^2}{\alpha_{\|}^2\alpha_{\bot}^2} \frac{1-\frac{1}{2}q_0^2r^2}{\left(1+q_0^2r^2\right)^2}
\nonumber\\[.1cm]
&&{}
\qquad-\frac{\zeta_3}{\alpha_{\bot}^2}\frac{1-\frac{1}{2}q_0^2r^2}{1+q_0^2r^2}-2\frac{\zeta_4\alpha_{\|}\alpha_{\bot}+\zeta_6\alpha_{\|}(\alpha_{\bot}-\alpha_{\|})}{\alpha_{\|}^2\alpha_{\bot}^2}\frac{1}{1+q_0^2r^2}-\frac{\zeta_6}{\alpha_{\bot}^2}
\bigg\}\mu_0^2H^2A
\nonumber\\
&&{}
-\bigg\{
\frac{\alpha_{\|}-\alpha_{\bot}}{\alpha_{\|}\alpha_{\bot}}
\frac{q_0^2r^2}{\left(1+q_0^2r^2\right)^2}
+\frac{1}{2}\frac{\zeta_1\alpha_{\bot}^2+\zeta_3(\alpha_{\bot}^2-\alpha_{\|}^2)+2\zeta_4\alpha_{\bot}({\alpha_{\bot}}-\alpha_{\|})+\zeta_6(\alpha_{\bot}-\alpha_{\|})^2}{\alpha_{\|}^2\alpha_{\bot}^2} \frac{q_0^2r^2}{\left(1+q_0^2r^2\right)^2}
\nonumber
\\[.1cm]
&&{}
\qquad+\frac{1}{2}\frac{\zeta_3}{\alpha_{\bot}^2}\frac{q_0^2r^2}{1+q_0^2r^2}
+\frac{1}{2}\frac{\zeta_4\alpha_{\|}\alpha_{\bot}+\zeta_6\alpha_{\|}(\alpha_{\bot}-\alpha_{\|})}{\alpha_{\|}^2\alpha_{\bot}^2}\frac{q_0^2r^2}{1+q_0^2r^2}
\bigg\}\mu_0^2H^2\frac{\tau}{q_0}.
\label{eq_macro_exp}
\end{eqnarray}
From here, the macroscopic free energy follows via $\mathcal{F}^{\,\mathrm{macro}}=\int \mathrm{d}^3r\,F^{\,\mathrm{macro}}$. An expression for the amplitude $A$ of elastic contraction and expansion is then obtained from $\partial\mathcal{F}^{\,\mathrm{macro}}/\partial A=0$ as
\begin{eqnarray}
A&=&
{}-\frac{1}{3\mu R^2}\bigg\{
\frac{\alpha_{\|}-\alpha_{\bot}}{\alpha_{\|}\alpha_{\bot}}
\left[\frac{3}{2q_0^2}\ln\left(1+q_0^2R^2\right)-\frac{3R^2}{2\left(1+q_0^2R^2\right)}\right]
\nonumber\\
&&{}
\qquad
-\frac{\zeta_1\alpha_{\bot}^2+\zeta_3(\alpha_{\bot}^2-\alpha_{\|}^2)+2\zeta_4\alpha_{\bot}({\alpha_{\bot}}-\alpha_{\|})+\zeta_6(\alpha_{\bot}-\alpha_{\|})^2}{\alpha_{\|}^2\alpha_{\bot}^2} \left[\frac{3}{4}\frac{R^2}{1+q_0^2R^2}-\frac{1}{4q_0^2}\ln\left(1+q_0^2R^2\right)\right]
\nonumber\\
&&{}
\qquad
+\frac{\zeta_3}{\alpha_{\bot}^2}
\left[\frac{1}{4}R^2-\frac{3}{4q_0^2}\ln\left(1+q_0^2R^2\right)\right]
-2\frac{\zeta_4\alpha_{\|}\alpha_{\bot}+\zeta_6\alpha_{\|}(\alpha_{\bot}-\alpha_{\|})}{\alpha_{\|}^2\alpha_{\bot}^2}
\frac{1}{2q_0^2}\ln\left(1+q_0^2R^2\right)
-\frac{\zeta_6}{\alpha_{\bot}^2}\frac{R^2}{2}
\bigg\}\mu_0^2H^2
.\qquad
\end{eqnarray}
Moreover, the amplitude $\tau$ of twist deformation resulting from $\partial\mathcal{F}^{\,\mathrm{macro}}/\partial\tau=0$ reads
\begin{eqnarray}
\tau &=&
\frac{4}{\mu\,q_0R^4}\bigg\{
\frac{\alpha_{\|}-\alpha_{\bot}}{\alpha_{\|}\alpha_{\bot}}
\left[\frac{1}{2q_0^2}\ln\left(1+q_0^2R^2\right)-\frac{R^2}{2\left(1+q_0^2R^2\right)}\right]
\nonumber\\
&&{}
+\frac{1}{2}\frac{\zeta_1\alpha_{\bot}^2+\zeta_3(\alpha_{\bot}^2-\alpha_{\|}^2)+2\zeta_4\alpha_{\bot}({\alpha_{\bot}}-\alpha_{\|})+\zeta_6(\alpha_{\bot}-\alpha_{\|})^2}{\alpha_{\|}^2\alpha_{\bot}^2} \left[\frac{1}{2q_0^2}\ln\left(1+q_0^2R^2\right)-\frac{R^2}{2\left(1+q_0^2R^2\right)}\right]
\nonumber\\
&&{}
\qquad+\frac{1}{2}\frac{\zeta_3}{\alpha_{\bot}^2}
\left[\frac{R^2}{2}-\frac{1}{2q_0^2}\ln\left(1+q_0^2R^2\right)\right]
+\frac{1}{2}\frac{\zeta_4\alpha_{\|}\alpha_{\bot}+\zeta_6\alpha_{\|}(\alpha_{\bot}-\alpha_{\|})}{\alpha_{\|}^2\alpha_{\bot}^2}\left[\frac{R^2}{2}-\frac{1}{2q_0^2}\ln\left(1+q_0^2R^2\right)\right]
\bigg\}\mu_0^2H^2
.\qquad
\end{eqnarray}

As for the components of the magnetization $\mathbf{M}$, we obtain
\begin{eqnarray}
M_{\varphi} &=& 
\frac{\alpha_{\bot}-\alpha_{\|}}{\alpha_{\|}\alpha_{\bot}}
\frac{q_0r}{1+q_0^2r^2}\,\mu_0H
\nonumber\\
&&{}
+\bigg[
\frac{\alpha_{\|}-\alpha_{\bot}}{\alpha_{\|}\alpha_{\bot}}
\frac{3q_0r\left(1-q_0^2r^2\right)}{2\left(1+q_0^2r^2\right)^2}
+\frac{\zeta_1\alpha_{\bot}^2+\zeta_3(\alpha_{\bot}^2-\alpha_{\|}^2)+2\zeta_4\alpha_{\bot}({\alpha_{\bot}}-\alpha_{\|})+\zeta_6(\alpha_{\bot}-\alpha_{\|})^2}{\alpha_{\|}^2\alpha_{\bot}^2} \frac{q_0r\left(1-\frac{1}{2}q_0^2r^2\right)}{\left(1+q_0^2r^2\right)^2}
\nonumber\\
&&{}
\qquad+\frac{1}{2}\frac{\zeta_4+\zeta_6}{\alpha_{\|}\alpha_{\bot}}\frac{q_0r}{1+q_0^2r^2}
-\frac{1}{2}\frac{\zeta_6}{\alpha_{\bot}^2}\frac{q_0r}{1+q_0^2r^2}
\bigg]\mu_0H\,A
\nonumber\\
&&{}
+\bigg[
\frac{\alpha_{\bot}-\alpha_{\|}}{\alpha_{\|}\alpha_{\bot}}
\frac{q_0r\left(1-q_0^2r^2\right)}{\left(1+q_0^2r^2\right)^2}
+\frac{\zeta_1\alpha_{\bot}^2+\zeta_3(\alpha_{\bot}^2-\alpha_{\|}^2)+2\zeta_4\alpha_{\bot}({\alpha_{\bot}}-\alpha_{\|})+\zeta_6(\alpha_{\bot}-\alpha_{\|})^2}{\alpha_{\|}^2\alpha_{\bot}^2} \frac{q_0^3r^3}{\left(1+q_0^2r^2\right)^2}
\nonumber\\
&&{}
\qquad+\frac{1}{2}\frac{\zeta_4+\zeta_6}{\alpha_{\|}\alpha_{\bot}}q_0r
\bigg]\mu_0H\,\frac{\tau}{q_0}
.
\label{eq_Mmacro_phi}
\end{eqnarray}
Interestingly and in analogy to Eq.~(\ref{eq_Mmeso_z}) for the mesoscopic consideration in Sec.~\ref{sec_mesomodel}, we further find via explicit calculation
\begin{equation}\label{eq_Mmacro_z}
M_z=
{}-\frac{2}{\mu_0H} \left( F^{\,\mathrm{macro}} -\frac{3}{2}\mu A^2 - \frac{1}{2}\mu\tau^2r^2 \right).
\end{equation}
\end{widetext}

\section{Scale-bridging links}
\label{sec_scalebridging}

In Secs.~\ref{sec_mesomodel} and \ref{sec_macroscopic} we have now derived expressions for the energy densities and magnetizations within our mesoscopic model and the macroscopic theory, respectively. To link the two different scales, we can compare the expression for the mesoscopic energy density $\mathcal{W}/\pi R^2G$ in Eq.~(\ref{eq_meso_exp}) to the averaged macroscopic free-energy density $\mathcal{F}^{\,\mathrm{macro}}/\pi R^2G$ obtained from Eq.~(\ref{eq_macro_exp}). The same applies for the $\varphi$-components of the magnetizations, i.e., $\mathcal{M}_{\varphi}^{\mathrm{meso}}$ in Eq.~(\ref{eq_Mmeso_phi}) and the corresponding volume-averaged macroscopic expression obtained from Eq.~(\ref{eq_Mmacro_phi}). Due to the relations identified for $\mathcal{M}_z^{\mathrm{meso}}$ in Eq.~(\ref{eq_Mmeso_z}) and for $M_z$ in Eq.~(\ref{eq_Mmacro_z}), the comparison between the $z$-components of the magnetizations does not lead to any additional information. 

In each case, we compare separately those parts of the mesoscopic and averaged macroscopic expressions to each other that neither depend on $A$ nor on $\tau$, that are linear in $A$, and that are linear in $\tau$. Overall, this leads to six equations. They allow us to determine the six macroscopic system parameters $\alpha_{\|}$, $\alpha_{\bot}$, $\zeta_1$, $\zeta_3$, $\zeta_4$, and $\zeta_6$ as functions of the mesoscopic model parameters. These expressions correspond to the searched-for scale-bridging links.

\begin{widetext}
The parameters $\alpha_{\|}$ and $\alpha_{\bot}$ are identified from the comparison between the mesoscopic and averaged macroscopic expressions to vanishing order in $A$ and $\tau$ as
\begin{eqnarray}
\alpha_{\|} &=&
\mu_0(\chi+3)^2\left[q_0R-\arctan(q_0R)\right]\sqrt{h_0^2-2\rho_0^2\left[\cos(q_0h_0)-1\right]}^5
\bigg/
\nonumber\\
&&
\quad 
4\pi n\chi a^3\bigg( \left[q_0R-\arctan(q_0R)\right]
\left\{ (\chi+3)\sqrt{h_0^2-2\rho_0^2\left[\cos(q_0h_0)-1\right]}^5
+4\chi a^3\left[h_0^2-\rho_0^2+\rho_0^2\cos(q_0h_0)\right] \right\}
\nonumber\\
&&
\qquad
{}+3\chi a^3\rho_0h_0\sin(q_0h_0)\left[q_0^2R^2-\ln\left(1+q_0^2R^2\right)\right]\bigg)
,
\label{eq_alphap}
\\
\alpha_{\bot} &=& 
\mu_0(\chi+3)^2\left[q_0R-\arctan(q_0R)\right]\sqrt{h_0^2-2\rho_0^2\left[\cos(q_0h_0)-1\right]}^5
\bigg/
\nonumber\\
&&
\quad 
4\pi n\chi a^3\bigg( \left[q_0R-\arctan(q_0R)\right]
\left\{ (\chi+3)\sqrt{h_0^2-2\rho_0^2\left[\cos(q_0h_0)-1\right]}^5
+4\chi a^3\left[h_0^2-\rho_0^2+\rho_0^2\cos(q_0h_0)\right] \right\}
\nonumber\\
&&
\qquad
{}-3\chi a^3\rho_0h_0\sin(q_0h_0)\ln\left(1+q_0^2R^2\right)\bigg)
.
\label{eq_alphas}
\end{eqnarray}
\end{widetext}
Expressions for the macroscopic system parameters $\zeta_1$, $\zeta_3$, $\zeta_4$, and $\zeta_6$ in terms of the mesoscopic model parameters are then calculated by comparing the remaining parts of the mesoscopic and averaged macroscopic relations linear in $A$ and in $\tau$ with each other. The procedure corresponds to solving a linear system of equations. Since the resulting expressions are very lengthy, we do not list them here explicitly. Instead, we plot the dependence of all six macroscopic system parameters on $q_0$, which quantifies the dependence on the initial twist of the particle structure. 

For this purpose, numerical values need to be assigned to the remaining mesoscopic model parameters. A first set of parameter values is inspired by the experimental system investigated in Ref.~\onlinecite{puljiz2018reversible}. There, the behavior of paramagnetic nickel particles of radius $a\approx85$~$\mu$m and relative magnetic permeability $\mu_{\mathrm{r}}\approx14.1$, implying $\chi\approx13.1$, was analyzed. 
The there-measured distance between these particles in the nonmagnetized state inspires our choice of $h_0\approx300$~$\mu$m. Typical cylindrical samples of magnetorheological elastomers show a radius, for example, of $R\approx2$~mm \cite{schumann2017situ}. Accordingly, we set $\rho_0\approx1.5$~mm. Considering $6$ initially twisted chain-like aggregates located at distance $\rho_0$ around the cylinder axis, the particle number density can be approximated as $n\approx10^{10}/2\pi$~m$^{-3}$. 

\begin{figure}
\includegraphics[width=8.3cm]{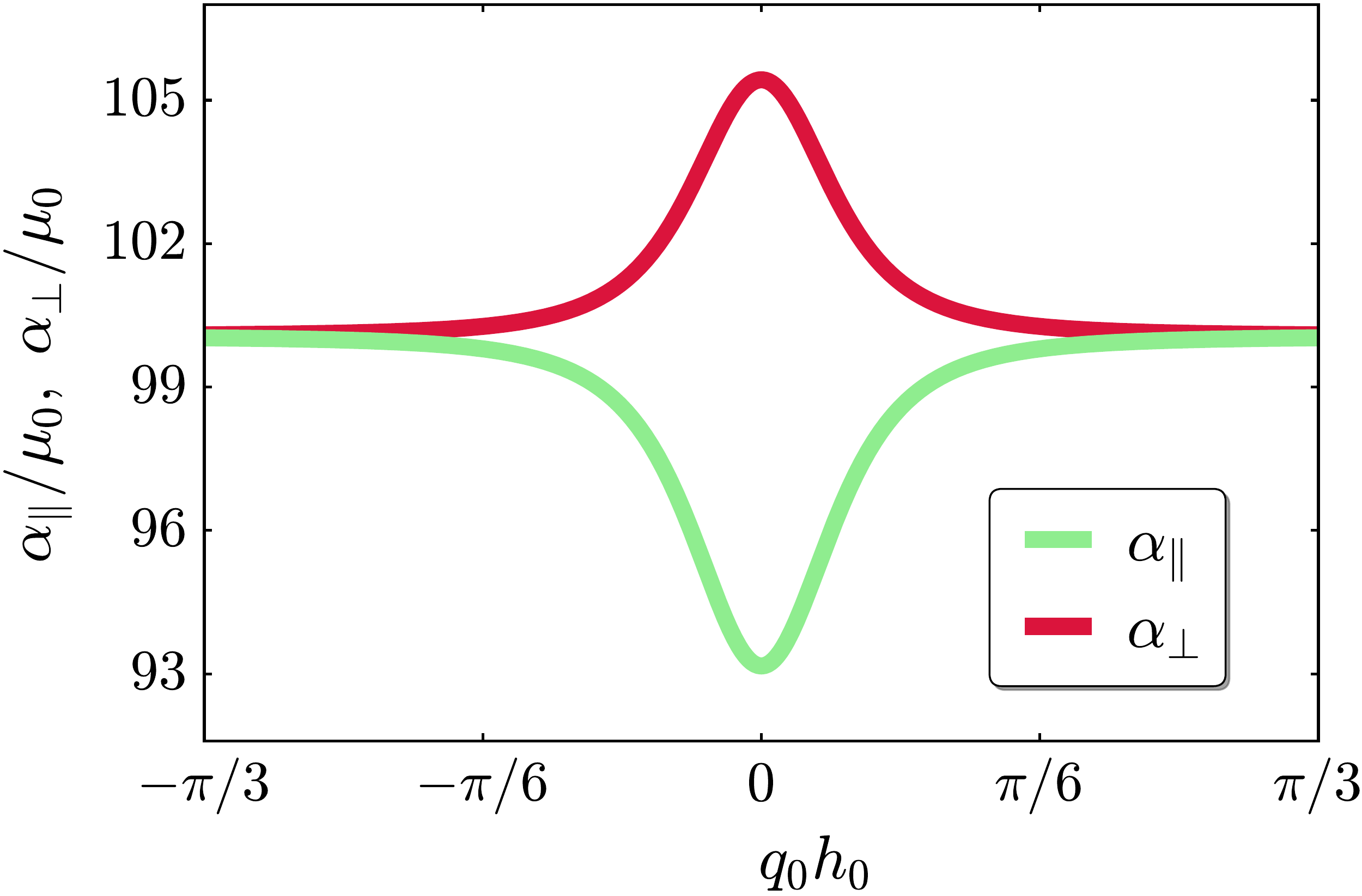}
\caption{Macroscopic system parameters $\alpha_{\|}$ and $\alpha_{\bot}$, see Eqs.~(\ref{eq_alphap}) and (\ref{eq_alphas}), rescaled by $\mu_0$, as functions of the scaled initial twist $q_0h_0$ and as calculated from the underlying mesoscopic model parameters. Here, $h_0$ is kept fixed and $q_0$ is varied. The numerical values of $h_0$ and of the other mesoscopic model parameters are kept constant as described in the main text. \am{An offset along the ordinate is imposed for illustration to stress the variations in the parameter values.}}
\label{fig_alphas}
\end{figure}

Along these lines, Fig.~\ref{fig_alphas} shows the dependence of the macroscopic system parameters $\alpha_{\|}$ and $\alpha_{\bot}$, see Eqs.~(\ref{eq_alphap}) and (\ref{eq_alphas}), on $q_0$. 
As expected, the corresponding curves are symmetric with respect to the axis $q_0=0$. From Eq.~(\ref{eq_F_macro}), it is obvious that the parameters $\alpha_{\|}$ and $\alpha_{\bot}$ by themselves play the roles of inverse magnetic susceptibilities, if we confine ourselves to the linear regime. It is then illustrative from the mesoscopic picture that for small values of $|q_0|$ the magnitude of $\alpha_{\|}$ is smallest (i.e., the corresponding magnetic susceptibility is largest) for $q_0=0$. In this case, as Fig.~\ref{fig_cylinder_meso} indicates, $\gamma_0=0$, and therefore the particles form vertical chain-like aggregates in the initial state. Their separation distance $d$ is then smallest, which supports their mutual magnetization. Conversely, for small values of $|q_0|$ the magnitude of $\alpha_{\bot}$ is largest (i.e., the corresponding magnetic susceptibility is smallest) for $q_0=0$. This can likewise be illustratively understood from the mesoscopic picture. When the chain-like aggregates are magnetized perpendicular to their axes, the mutual magnetization between neighboring particles counteracts their net magnetization. The closed magnetic induction field lines running through one particle due to its own magnetization point into the opposite direction at the location of the neighboring particles. As before, we observe the strongest effect when the distance $d$ between the particles is smallest, i.e., for $q_0=0$. Overall, we find $\alpha_{\|}<\alpha_{\bot}$ for small $|q_0|$, which implies a larger magnetic susceptibility along the chain-like aggregates than perpendicular to them, as expected. 

\begin{figure}
\includegraphics[width=8.3cm]{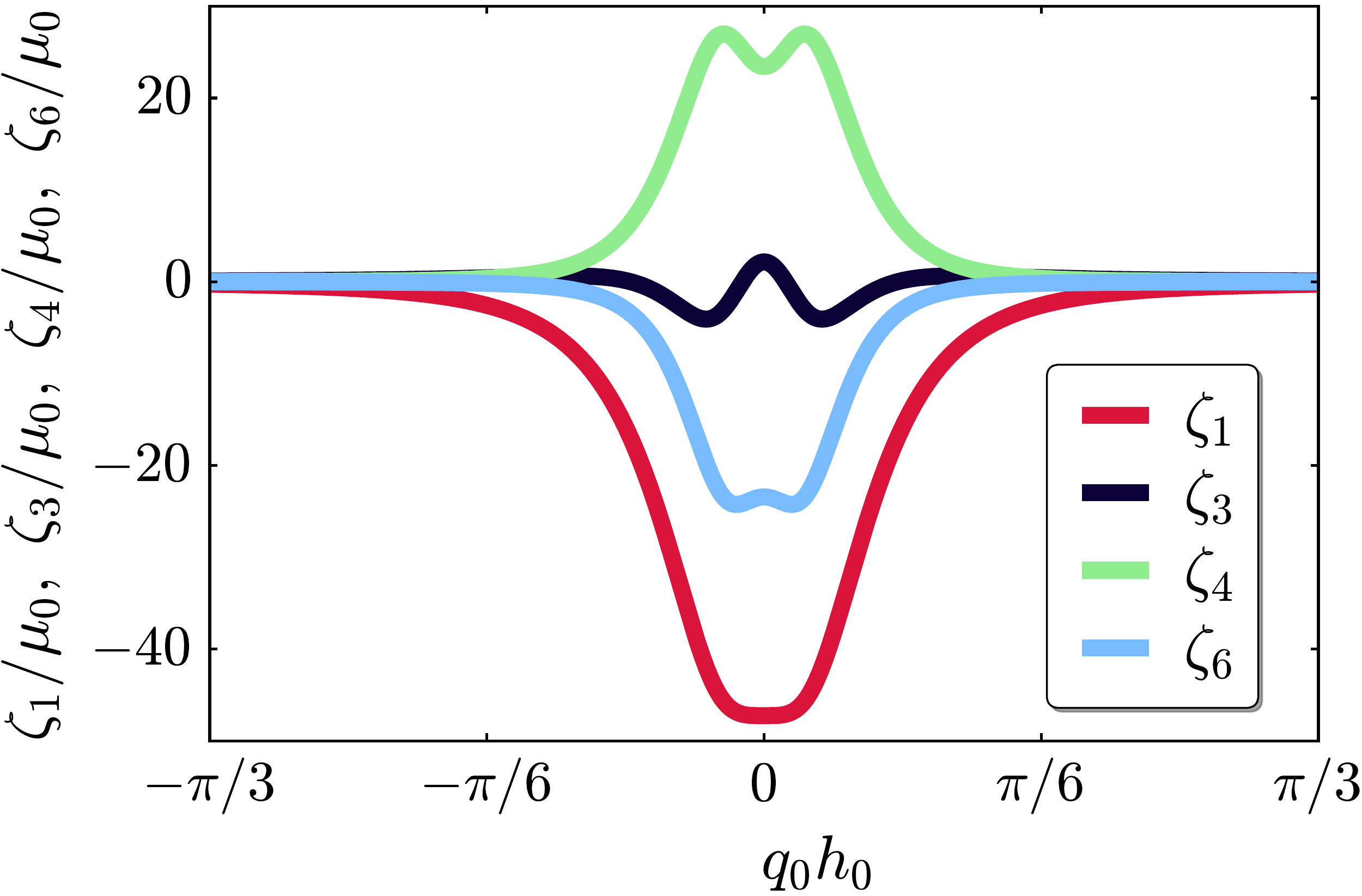}
\caption{Macroscopic system parameters $\zeta_1$, $\zeta_3$, $\zeta_4$, and $\zeta_6$, rescaled by $\mu_0$, as functions of the scaled initial twist $q_0h_0$ and as calculated from the underlying mesoscopic model parameters. Again, $q_0$ is varied, while the numerical values of all other mesoscopic model parameters are kept constant as listed in the main text.}
\label{fig_gammas}
\end{figure}

Analogous results for the macroscopic system parameters $\zeta_1$, $\zeta_3$, $\zeta_4$, and $\zeta_6$ as functions of $q_0$ and as calculated from the mesoscopic model parameters are shown in Fig.~\ref{fig_gammas}. 
Since the explicit analytical expressions are relatively complex, the numerical evaluation near $q_0=0$ is quite involved and it is possible that Fig.~\ref{fig_gammas} contains some deviations in this region. 
Nevertheless, the spacial proximity between the particles in the mesoscopic model again implies a most pronounced effect for not too large $|q_0|$. This favors the mutual magnetic particle interactions and leads to the elevated magnitudes of $\zeta_1$, $\zeta_3$, $\zeta_4$, and $\zeta_6$ in this region. In our case, $\zeta_1<0$ implies that a magnetic field along the local structural axis $\mathbf{\hat{n}}$ induces a contraction along this axis, see Eqs.~(\ref{eq_Fmstr}) and (\ref{eq_zeta}). $\zeta_3<0$ implies contractions along $\mathbf{\hat{n}}$ independently of the direction of magnetization, while expansions occur for $\zeta_3>0$. As we can see from Fig.~\ref{fig_gammas}, the magnitude of $\zeta_3$ at small $|q_0|$ is relatively small when compared to the other system parameters. For oblique orientations of $\mathbf{M}$ relative to $\mathbf{\hat{n}}$, $\zeta_4\neq0$ is related to magnetically induced shear deformations within the plane spanned by $\mathbf{M}$ and $\mathbf{\hat{n}}$. Finally, $\zeta_6<0$ implies magnetically induced contractions along $\mathbf{M}$. 
\am{The nonmonotonous behavior of the parameters in this regime as a function of $q_0$ becomes plausible when we remind ourselves that both linear-type and torsional deformations contribute to the magnetostrictive response. The torsional part in our geometry vanishes for $q_0=0$.}

Concerning the interpretation of the results for larger values of $|q_0|$, we need to recall the different constructions of the mesoscopic and macroscopic descriptions. Within the mesoscopic minimal model, where here we assumed six chain-like aggregates, $q_0h_0=\pi/3$ implies that the structure is initially twisted from one particle to the next one to an amount that positions the first particle vertically below the next particle of a neighboring chain. In reality, this situation would be identical to the one for $q_0=0$. It  identifies straight vertical undeformed chain-like particle aggregates. In contrast to that, the tilt of $\mathbf{\hat{n}}$ relative to the vertical direction in the macroscopic theory by definition simply keeps monotonically increasing with increasing $|q_0|$. 

Moreover, as just indicated, with increasing $|q_0|$, particles of what we consider the same chain in the mesoscopic picture can be located further apart from each other than particles belonging to different chains. Then, our approximation of only considering magnetic interactions between neighboring particles on the same chain becomes questionable. This needs to be kept in mind for larger values of $|q_0|$. Still the mesoscopic model picture could be used to describe the phenomenology observed for a corresponding macroscopic system in many cases. It is only that the mesoscopic model parameters need to be adjusted accordingly to lead to the correct macroscopic phenomenology when the macroscopic system parameters are calculated as outlined above. In reality, if the samples are fabricated by a two-step crosslinking process as outlined in Sec.~\ref{sec_introduction}, the implementation of such large values of $|q_0|$ in actual samples appears questionable after all, which argues for our approach.

\section{An effective macroscopic phenomenological picture}
\label{sec_macrocomplete}

Altogether, we have now addressed the answer to our question asked at the end of Sec.~\ref{sec_macroproblem} --- how do we have to formulate our picture to include the effect of magnetically induced twist deformations --- from various sides. As we have seen, in parts the underlying formulae become relatively lengthy.

Possibly, quantities that are measured in a macroscopic experiment may be the overall elastic deformations quantified by $A$ and $\tau$, perhaps supplemented by the overall magnetic moment $\bm{\mathcal{M}}=\int\mathrm{d}^3r\,\mathbf{M}=\mathcal{M}_z\mathbf{\hat{z}}$ of the whole system. By construction, this set of three quantities corresponds to a restricted point of view on the whole system behavior, as many degrees of freedom are not taken into account. 

However, it comes as a positive surprise that the associated restricted macroscopic phenomenology corresponding to this point of view can already be described qualitatively by ignoring the magnetostrictive terms, that is, by setting $\zeta_1=\zeta_3=\zeta_4=\zeta_6=0$. We stress that the magnetostrictive terms may well be of identical order as the other coupling terms containing analogous effects. It is only the restricted point of view that does not allow us to distinguish between the actual origin of the observed phenomenology any longer. 

Accordingly, we obtain from Eq.~(\ref{eq_macro_exp})
\begin{eqnarray}
\frac{\mathcal{F}^{\,\mathrm{macro}}}{\pi R^2G} &=& \frac{3}{2}\mu A^2 + \frac{1}{4}\mu R^2\tau^2
\nonumber\\
&&{}
+\xi_1H^2+3\xi_2H^2A-2\xi_2H^2\frac{\tau}{q_0},
\end{eqnarray}
where we have defined
\begin{eqnarray}
\xi_1 &=&
\frac{\mu_0^2}{2\alpha_{\bot}}\left[ \frac{\alpha_{\|}\!-\!\alpha_{\bot}}{\alpha_{\|}}\frac{1}{q_0^2R^2}\ln\left(1+q_0^2R^2\right)-1 
\right],
\\[.1cm]
\xi_2 &=& \mu_0^2\,\frac{\alpha_{\|}\!-\!\alpha_{\bot}}{2\alpha_{\|}\alpha_{\bot}}\left[
\frac{1}{q_0^2R^2}\ln\left(1+q_0^2R^2\right)-\frac{1}{1+q_0^2R^2}
\right].
\nonumber\\
&&
\end{eqnarray}
Minimizing $\mathcal{F}^{\,\mathrm{macro}}$ with respect to $A$ and $\tau$, we find 
\begin{eqnarray}
A&=&{}-\frac{1}{\mu}\,\xi_2\,H^2, \\
\tau&=&\frac{4\xi_2}{\mu R^2}\frac{1}{q_0}H^2.
\end{eqnarray}
Therefore, plotting $A(H^2)$, we can determine $\xi_2$. Additionally plotting in a second step $\tau(H^2)$, we can identify $q_0$. In combination, $(\alpha_{\|}-\alpha_{\bot})/\alpha_{\|}\alpha_{\bot}$ follows from $\xi_2$ and $q_0$. 

We stress that this restricted phenomenological point of view turns $q_0$ into a fit parameter, in contrast to our deterministic point of view taken in Secs.~\ref{sec_mesomodel}--\ref{sec_scalebridging}. In fact, without detailed knowledge on the actual processes occurring during fabrication and during the implementation of the initial global twist of the sample on the mesoscopic scale, such an approach may be reasonable in reality. It implies that details on the resulting internal mesoscopic structure are not available or are simply not taken into account. 

If, additionally, the overall magnetic moment $\bm{\mathcal{M}}=\int\mathrm{d}^3r\,\mathbf{M}=\mathcal{M}_z\mathbf{\hat{z}}$ of the whole system can be measured as a function of $H$, then $\alpha_{\|}$ and $\alpha_{\bot}$ can be determined. More precisely, via Eq.~(\ref{eq_Mmacro_z}) we calculate
\begin{eqnarray}\label{eq_curlyMz}
\mathcal{M}_z&=&\int\mathrm{d}^3r\,M_z
\nonumber\\
&=&
\pi G\bigg(
\frac{\mu_0}{\alpha_{\bot}}R^2
-\mu_0\frac{\alpha_{\|}-\alpha_{\bot}}{\alpha_{\|}\alpha_{\bot}} \frac{1}{q_0^2}\ln\left(1+q_0^2R^2\right)
\nonumber\\[.1cm]
&&{}
+\frac{6\xi_2^2}{\mu_0\,\mu}R^2H^2
+\frac{16\xi_2^2}{\mu_0\,\mu}\frac{1}{q_0^2}H^2
\bigg)H.
\end{eqnarray}
$\xi_2$ and $(\alpha_{\|}-\alpha_{\bot})/\alpha_{\|}\alpha_{\bot}$ are already known from the previous plots. Additionally plotting $\mathcal{M}_z(H)$, we can thus determine $\alpha_{\bot}$ via Eq.~(\ref{eq_curlyMz}) and consequently $\alpha_{\|}$ from $(\alpha_{\|}-\alpha_{\bot})/\alpha_{\|}\alpha_{\bot}$. 

What is the benefit of this procedure? Having determined from $A$, $\tau$, and $\mathcal{M}_z$ the system parameters $q_0$, $\alpha_{\|}$, and $\alpha_{\bot}$, we can via our scale-bridging links introduce a corresponding effective mesoscopic model system as in Figs.~\ref{fig_geometry} and \ref{fig_cylinder_meso}. The mesoscopic model parameters need to be adjusted accordingly to result in identical values of the macroscopic system parameters as extracted from corresponding experiments. Frequently, several aspects can be more illustratively understood on such a mesoscopic model system than from the global behavior of an actual macroscopic sample. For instance, this may be because of profound polydispersity of the magnetizable particles or because of much more irregular particle structures in the real system. We wish to emphasize once more, however, that the mapping as just described as a reduction of our approach in Secs.~\ref{sec_mesomodel}--\ref{sec_scalebridging} only applies within our restricted phenomenology that solely concentrates on $A$, $\tau$, and $\mathcal{M}_z$.

\section{Conclusions}
\label{sec_conclusions}

In summary, we have described on a discrete mesoscopic and on a macroscopic level the magnetically induced elastostatic deformation of a soft twist actuator based on magnetorheological gels or elastomers. The construction of such a device has recently been suggested \cite{fischer2020towards}. Comparing between the mesoscopic and macroscopic levels allows us to extract explicit expressions for the macroscopic system parameters as functions of the mesoscopic model parameters. In this way, a bridge between these two scales is established. The mesoscopic model guided us to the corresponding macroscopic approach. First, we considered a macroscopic picture including magnetostrictive contributions to the free-energy density. Afterwards, we simplified the description by neglecting such contributions in the case of a reduced macroscopic phenomenology. The latter approach may potentially correspond to some common experimental measurements. 

On our way, we approached the mesoscopic scale by a genuine minimal model. Such a model that correctly represents the macroscopic phenomenology can significantly support the illustrative understanding of the underlying physical effects. Nevertheless, in the future, more refined descriptions could be established also on the mesoscopic level. This includes less regular particle arrangements, nonaffine deformations, and imperfections in the particle shapes, sizes, or in the elasticity of the polymeric body. 
\am{For example, extended mesoscopic models that address twisted structures of particles of binary size distribution have already been evaluated numerically, taking into account nonaffine deformations \cite{fischer2020magnetically}.}

\am{Overall, we hope that our study can further stimulate the experimental realization of soft twist actuators based on magnetorheological gels and elastomers. At the beginning, we have outlined ideas for possible routes of fabrication. One way to quantify the experimentally induced deformations could be to immerse the device in an optically transparent fluid\cite{gollwitzer2008measuring} and to add optically detectable tracer lines to its surfaces. Magnetically induced deformations then become visible by the distorted lines. If information on the internal structure and particle arrangement is available, for example, using x-ray microtomography \cite{schumann2019microscopic}, the links of scale connection could be established on the experimental side for comparison with the theory. Possibly, the overall distortion may be inferred directly from x-ray microtomographic analyses as well, if both nonmagnetized and magnetized states of the whole device can be recorded.} 

If a corresponding actual device is used as a magnetic torsional actuator, for instance, in microfluidic mixing applications, analyzing the associated overall dynamics to optimize its performance will be another important aspect. 
\am{Correspondingly, the next step will be to also link theoretical approaches to the dynamics on the two scales. For this purpose, we first need to identify an appropriate mesoscopic model. Initially, it may be reasonable to start from overdamped dynamics of particle displacements\cite{puljiz2019memory}. Additionally, we then match expressions for the dissipation of energy on the different scales.}

We conclude by remarking that our approach is not restricted to the material class of magnetorheological gels and elastomers. For example, it equally applies to corresponding systems fabricated from electrorheological gels and elastomers \cite{an2003actuating,allahyarov2015simulation, liu2001electrorheology}. The mapping applies as long as electric leakage currents and associated dynamic effects upon electrically polarizing the systems are negligible.

\begin{acknowledgments}
The author appreciates stimulating discussions with Dr.\ G\"unter K.~Auernhammer and Professor Stefan Odenbach. Moreover, the author thanks the Deutsche Forschungsgemeinschaft (DFG, German Research Foundation) for support of this work through the Heisenberg grant ME 3571/4-1. 
\end{acknowledgments}

\section*{AIP publishing data sharing policy}

The data that support the findings of this study are available within the article. Figures \ref{fig_alphas} and \ref{fig_gammas} represent the direct numerical solutions of the derived sets of equations.

\appendix

\section{Changes of the geometric mesoscopic model parameters under elastic deformations to linear order}
\label{app_quantities}

In the following, we further elucidate the changes in the mesoscopic model parameters $h$, $\rho$, and $\gamma$, see also Fig.~\ref{fig_cylinder_meso}, as listed by Eqs.~(\ref{eq_h})--(\ref{eq_gamma}). Our expressions are derived to linear order in the elastic deformations as identified by the amplitudes $A$ and $\tau$. 

For simplicity, we assume that in Cartesian coordinates the position vector $\mathbf{r}_i$ of the $i$th particle in the initial, undeformed state of the system points into the direction $\mathbf{\hat{x}}$. Accordingly, 
\begin{equation}
\mathbf{r}_i=\left(\begin{array}{c}\rho_0\\[.1cm]0\\[.1cm]0\end{array}\right). 
\end{equation}
Consequently, for the $(i+1)$th particle, the position vector in the initial undeformed state is given by 
\begin{equation}
\mathbf{r}_{i+1}=\left(\begin{array}{c}\rho_0\cos\gamma_0\\[.1cm]\rho_0\sin\gamma_0\\[.1cm]h_0\end{array}\right). 
\end{equation}
As given by Eq.~(\ref{eq_q0}), the parameters $\gamma_0$ and $h_0$ are related via $\gamma_0=q_0h_0$. 

Next, evaluating the displacement field $\mathbf{u}$ identified in Eq.~(\ref{eq_u}) at the positions of these two particles, we calculate the corresponding two position vectors when the system is deformed. We obtain 
\begin{equation}\label{eq_riprime}
\mathbf{r}_i'=\mathbf{r}_i+\mathbf{u}_i=
\left(\begin{array}{c}\rho_0\left(1-\frac{1}{2}A\right)\\[.1cm]0\\[.1cm]0\end{array}\right) 
\end{equation}
and
\begin{eqnarray}
\mathbf{r}_{i+1}' &=& \mathbf{r}_{i+1}+\mathbf{u}_{i+1}
\nonumber\\[.1cm]
&=&\left(\begin{array}{c}\rho_0\cos\gamma_0\left(1-\frac{1}{2}A\right)-\tau\rho_0h_0\sin\gamma_0 
\\[.1cm]
\rho_0\sin\gamma_0\left(1-\frac{1}{2}A\right)+\tau\rho_0h_0\cos\gamma_0 
\\[.1cm]
h_0(1+A)\end{array}\right). \qquad
\label{eq_rip1prime}
\end{eqnarray}
The resulting changed distance $\rho$ between the particles and the center axis of the cylinder as given by Eq.~(\ref{eq_rho}) can be read off directly from the only nonvanishing component of $\mathbf{r}_i'$ in Eq.~(\ref{eq_riprime}). To linear order in $A$ and $\tau$, the same expression follows from the Cartesian $x$- and $y$-components of $\mathbf{r}_{i+1}'$ in Eq.~(\ref{eq_rip1prime}). Moreover, the changed vertical separation distance $h$ as given by Eq.~(\ref{eq_h}) can be read off from the $z$-component of $\mathbf{r}_{i+1}'$ in Eq.~(\ref{eq_rip1prime}). 

Finally, since $\mathbf{r}_i'\parallel\mathbf{\hat{x}}$, confining ourselves to the linear order in $A$ as well as in $\tau$, and using Eq.~(\ref{eq_q0}), we find
\begin{eqnarray}
\sin\gamma&=&\frac{y_{i+1}'}{\rho}
\nonumber\\[.1cm]
&= & \left[\sin\gamma_0\left(1-\frac{1}{2}A\right)+\tau h_0\cos\gamma_0 \right]\left(1+\frac{1}{2}A\right)
\nonumber\\[.1cm]
&= & \sin\gamma_0-\frac{1}{2}A\sin\gamma_0+\tau h_0\cos\gamma_0 +\frac{1}{2}A\sin\gamma_0
\nonumber\\[.1cm]
&=& \sin\gamma_0+\tau h_0\cos\gamma_0 
\nonumber\\[.15cm]
&= & \sin(\gamma_0+\tau h_0)=\sin\left[(q_0+\tau)h_0\right]. 
\label{eq_calc_gamma}
\end{eqnarray}
In this way, comparing the left- and right-hand sides of Eq.~(\ref{eq_calc_gamma}), we recover Eq.~(\ref{eq_gamma}).

\section{External magnetic field in the macroscopic free-energy density}
\label{app_a}

In general, the change in magnetic energy density can be denoted as \cite{jackson1962classical}
\begin{equation}
\mathrm{d} w=\mathbf{H}\cdot\mathrm{d}\mathbf{B},
\end{equation}  
where $\mathbf{B}$ is the magnetic induction field. From a thermodynamic point of view, when an external magnetic field $\mathbf{H}$ is imposed on the system and is kept fixed in an experiment, e.g., by an external electric current running through an external electric coil, a parameterization in terms of $\mathbf{H}$ appears more reasonable. For this purpose, a Legendre transformation is applied. 

First, we decompose $\mathbf{B}=\mu_0(\mathbf{H}+\mathbf{M})$. Thus, 
\begin{eqnarray}
\mathrm{d} w&=&\mu_0\,\left(\mathbf{H}\cdot\mathrm{d}\mathbf{H}+\mathbf{H}\cdot\mathrm{d}\mathbf{M}\right)
\nonumber\\
&=&\mu_0\,\left(\mathrm{d}\left(\frac{1}{2}\mathbf{H}^2\right)+\mathbf{H}\cdot\mathrm{d}\mathbf{M}\right).
\end{eqnarray}
Next, we define $\tilde{w}=w-\mu_0\,\mathbf{H}^2/2$, so that 
\begin{equation}\label{app_w}
\mathrm{d} \tilde{w}=\mu_0\,\mathbf{H}\cdot\mathrm{d}\mathbf{M}.
\end{equation}
From here, we apply a Legendre transformation 
\begin{equation}
\tilde{w}'=\tilde{w}-\mu_0\,\mathbf{H}\cdot\mathbf{M}
\end{equation}
to obtain
\begin{equation}
\mathrm{d} \tilde{w}'={}-\mu_0\,\mathbf{M}\cdot\mathrm{d}\mathbf{H}.
\end{equation}

Using in the first step Eq.~(\ref{app_w}), we find 
\begin{equation}
\frac{1}{\mu_0}\frac{\partial\tilde{w}}{\partial\mathbf{M}}=\mathbf{H}=\frac{\partial(\mathbf{H}\cdot\mathbf{M})}{\partial\mathbf{M}}.
\end{equation}
The outer parts on this line imply
\begin{equation}
\frac{\partial}{\partial\mathbf{M}}\left(\tilde{w}-\mu_0\,\mathbf{H}\cdot\mathbf{M}\right)=\frac{\partial\tilde{w}'}{\partial\mathbf{M}}=\mathbf{0}.
\end{equation}
Thus, under the given external magnetic field $\mathbf{H}$, we may use the free-energy density containing the subtracted term $\mu_0\,\mathbf{H}\cdot\mathbf{M}$ for a minimization with respect to $\mathbf{M}$ to find the actual state of the system.


%
%

%

\end{document}